# Near-field imaging beyond the probe aperture limit

Eunsung Seo[1,2,+], Young-Ho Jin[3,+], Wonjun Choi[1,2], Yonghyeon Jo[1,2], Suyeon Lee[4], Kyung-Deok Song[1,2], Joonmo Ahn[1,2], Q-Han Park[2], Myung-Ki Kim[3,*], Wonshik Choi[1,2,*]

[1]Center for Molecular Spectroscopy and Dynamics, Institute for Basic Science, Seoul 02841, Korea
[2]Department of Physics, Korea University, Seoul 02841, Korea
[3]KU-KIST Graduate School of Converging Science and Technology, Korea University, Seoul 02841, Korea
[4]Samsung Advanced Institute of Technology, 130, Samsung-ro, Yeongtong-gu, Suwon, Gyeonggi-do, 16678, Korea
[+]These authors contributed equally to this work.
[*]e-mail: rokmk@korea.ac.kr and wonshik@korea.ac.kr

**Abstract:**

Near-field scanning optical microscopy has been an indispensable tool for designing, characterizing and understanding the functionalities of diverse nanoscale photonic devices[1-4]. As the advances in fabrication technology have driven the devices smaller and smaller, the demand has grown steadily for improving its resolving power, which is determined mainly by the size of the probe attached to the scanner[5]. The use of a smaller probe has been a straightforward approach to increase the resolving power, but it cannot be made arbitrarily small in practice due to the steep reduction of the collection efficiency. Here, we develop a method to enhance the resolving power of near-field imaging beyond the limit set by the physical size of the probe aperture. The main working principle is to unveil high-order near-field eigenmodes invisible with conventional near-field microscopy. The destructive interference of near-field waves is induced in these high-order eigenmodes by the locally varying phases, which can reveal subaperture-scale fine structural details. To extract these eigenmodes, we construct a self-interference near-field microscopy system and measure a fully phase-referenced far-to near-field transmission matrix (FNTM) composed of near-field amplitude and phase maps recorded for various angles of far-field illumination[6,7]. By the singular value decomposition of the measured FNTM, we could extract the antisymmetric mode[8,9], quadrupole mode, and other higher-order modes hidden under the lowest-order symmetric mode. This enables us to resolve double and triple nano-slots whose gap size (50 nm) is three times smaller than the diameter of the probe aperture (150 nm). The subaperture near-field mode mapping by the FTNM can be potentially combined with various existing near-field imaging modalities[1,10-13] and promote their ability to interrogate local near-field optical waves of nanoscale devices.

**Main text:**

Enhancing the spatial resolving power beyond the diffraction limit has always been a fascinating subject in the field of optical imaging since it allows us to reveal fine structural details invisible with the conventional imaging modalities. Due to this strong demand, various forms of super-resolution microscopy have been proposed over the past decades based on the diverse working mechanisms[14-18]. The explicit approach to overcome the diffraction limit requires the exploitation of the near-field evanescent waves carrying spatial frequencies exceeding the far-field limit set by $k_0 = 2\pi/\lambda$, where $\lambda$ is the wavelength of light source[5]. Near-field scanning optical microscopy (NSOM) achieves this by using subwavelength probes, which convert non-propagating near-field waves to far-field propagating waves[19]. By using a probe much smaller than the diffraction limit, the spatial resolution of NSOM reaches tens of nanometers. NSOM has been applied to various fields of research such as plasmonic/photonic mode analysis[4,10,20], Raman spectroscopy[21], and single-molecule detection[22].

In recent years, the demand for improving the resolving power of NSOM has steadily grown owing to the relentless advancement of nanofabrication techniques. However, as far-field imaging is diffraction-limited, the spatial resolution of NSOM meets a hard limit set by the physical size of the probe. A straightforward approach to enhance spatial resolution has been to reduce the probe size. However, the probe cannot be made arbitrarily small due to the significant reduction in collection efficiency. For example, an aperture-type NSOM using a metal-coated tapered fiber as a probe typically sets the aperture diameter to around 100 nm because the collection efficiency decreases below $10^{-6}$ at aperture sizes smaller than 50 nm and decays faster than even exponentially with the decrease in the aperture size[5]. Similarly, the size of the probe used in the scattering-type NSOM system is limited by the strength of the Rayleigh scattering at the probe tip, which decreases in proportion to the sixth power of its size[23]. Field enhancement between the probe tip and the sample is often used to increase the signal strength; however, a careful analysis should be conducted as the strong tip-sample interaction often obscures the intrinsic near-field distribution of the sample[24].

Here, we propose a near-field imaging method, which resolves near-field eigenmodes beyond the limit set by the physical size of the probe aperture. For the nano-slots whose gap size is significantly smaller than the diameter of the probe aperture, we develop a method to map the high-order near-field eigenmodes whose spatial variations are too fine to detect by the conventional NSOM. High-order eigenmodes are barely excited by the far-field coupling to nanostructures whose spatial variations are finer than the wavelength, which explains the reason why only the symmetric modes are visible in the conventional NSOM. For the proposed method, we construct an interferometric NSOM system, which measures a fully phase-referenced far- to near-field transmission matrix (FNTM), and identify the near-field eigenmodes by the singular value decomposition (SVD) of the measured matrix. In this way, we extract weak high-order modes obscured by the strong symmetric mode with greatly improved sensitivity. Among various high-order eigenmodes, we find that the antisymmetric mode is particularly useful for the increase in resolving power. In the antisymmetric mode, a steep phase jump exists in the middle of the two neighboring nano-slots as the radiations at the two slots are out of phase. By taking the advantage of this phase gradient, we can clearly resolve the two slots whose gap size is as small as 50 nm using a probe aperture having a diameter of 150 nm, three times larger than the gap size. As a generalization of the proposed method, we resolve triple nano-slots by mapping the second-order transverse eigenmodes. In addition to the antisymmetric mode, even higher-order near-field modes such as quadrupole and third-order modes are revealed, which carry rich information on the light interaction with nanostructures.

The mapping of the antisymmetric mode provides subaperture resolving power as it breaks the condition where the diffraction limit is originally defined. In the conventional far-field imaging, two incoherent point sources whose relative phase is random are considered, where the average of the intensities by the point sources smoothens the structural details. In the antisymmetric mode, strong destructive interference between the emitters causes a steep spatial variation in the net phase in the middle of the emitters[25]. This feature is preserved even after the convoluted detection by an aperture larger than the slot separation, which is the main mechanism of our subaperture NSOM imaging. It is worth noting that super-resolution fluorescence microscopy techniques could overcome the diffraction limit even though only far-field optics is used. This is because they also alleviated the conventional setting where the diffraction limit is defined. Wide-field approaches rely on the sequential switching of fluorescence molecules[15,16,26], and the stimulated emission depletion approach makes use of the strong nonlinear saturation effect[27]. However, the excitation of the antisymmetric mode has been challenging as the out-of-phase excitation of two emitters located within the subwavelength aperture is hard to realize by the far-field illumination. Usually, the emitters are driven nearly in phase so that the conventional NSOM could observe only the symmetric mode. As discussed below, our experimental approach of measuring the FNTM is extremely sensitive to reveal various high-order near-field modes including the antisymmetric mode hidden under strong lowest-order symmetric modes.

A transmission matrix describes the coherent linear interaction between light and arbitrary device including disordered media. It describes the complex-field map, i.e., phase and amplitude maps, at the output plane of the device depending on the excitation of each individual mode at the input plane. It has been widely used in the far-field regime in the past and offered unique opportunities. For example, the inversion of the transmission matrix led to image delivery through a scattering layer[6], while the eigenmodes of the transmission matrix were used for an efficient light energy delivery and mapping target objects within scattering medium[6,28-31]. The transmission matrix approach also allows for the control and image delivery of near-field waves by exploiting plasmonic waves and near-field waves generated by the disordered medium[32-35]. In this study, we developed an experimental method to record a phase-referenced FNTM, $t(x, y; \vec{k}^{in})$, which describes the near-field complex-field maps at the upper surface $(x, y, z = 0)$ of the nanostructures for the far-field illumination at the bottom surface $(z = -z_0)$ with various transverse wavevectors, $\vec{k}^{in} = (k_x^{in}, k_y^{in})$. Figure 1 presents a simplified experimental scheme where a planar far-field incident wave with a given $\vec{k}^{in}$ was sent from the bottom of nanostructures, and the subwavelength-scale aperture probe mounted on NSOM picked up the near-field wave at the upper surface. The important feature of our method is the recording of both amplitude and phase of the near-field wave by using the self-interference phase-shifting interferometry. To this end, we installed a spatial light modulator (SLM) at the conjugate plane to the sample plane in the illumination beam path and wrote a phase pattern on the SLM to generate two plane waves at the bottom of the nano-slots, one with normal incidence angle and the other with $\vec{k}^{in}$:

$$E_{in}(x, y, z = -z_0; \vec{k}^{in}, \Delta\phi_R) = A_0 e^{-i\Delta\phi_R} + A_0 e^{-i(k_x^{in}x + k_y^{in}y)}. \qquad (1)$$

Here, the normally incident plane wave, whose wavefronts are indicated by the red lines in Fig. 1a, serves as a reference wave. Its relative phase $\Delta\phi_R$ with respect to the other sample wave (indicated by the blue lines in Fig. 1a) was controlled by the phase pattern written on the SLM. The transmitted wave on the upper surface of the nanostructures can be expressed as:

$$E_{out}(x, y, 0; \vec{k}^{in}, \Delta\phi_R) = E_R(x, y)e^{-i\Delta\phi_R} + E_S(x, y; \vec{k}^{in}), \qquad (2)$$

where $E_R(x,y)$ and $E_S(x,y;\vec{k}^{in})$ are the complex-field amplitudes of the reference and sample waves, respectively, at the upper surface. The NSOM probe recorded the interference intensity $I_{out} = |E_{out}(x,y,0;\vec{k}^{in},\Delta\phi_R)|^2$, which is expressed as a sinusoidal function of $\Delta\phi_R$. By measuring the intensities at four incremental steps of $\Delta\phi_R$, i.e., $\Delta\phi_R = 0, \frac{\pi}{2}, \pi$ and $\frac{3\pi}{2}$, we could demodulate the amplitude and phase of $E_S(x,y;\vec{k}^{in})$ [36]. Here, we accounted for the amplitude of $E_R(x,y)$ by separately measuring its intensity only for the normal illumination. The phase of $E_R(x,y)$ was assumed to be flat, which is typical for the symmetrically driven subwavelength nanostructures. In the experiment, we chose an illumination wavevector $\vec{k}^{in}$ and sequentially displayed four phase patterns on the SLM setting $\Delta\phi_R$ to multiples of $\frac{\pi}{2}$. The NSOM probe was scanned across the lateral plane at the upper surface to obtain four corresponding near-field intensity maps. By applying the phase-shifting interferometry algorithm, we obtained the near-field complex-field map for the corresponding $\vec{k}^{in}$, i.e. $E_S(x,y;\vec{k}^{in})$, and used it to construct the FNTM.

To validate the proposed method, we prepared a pair of nano-slots whose eigenmodes are theoretically well known. On a 100-nm-thick gold film, we fabricated double nano-slots using a focused ion beam milling (bottom image in Fig. 1b) (see Methods for the sample fabrication). The width $W$ of each slot was approximately 20 nm, which sets the effective refractive index of the plasmonic modes within the slot to approximately 2.0. The length $L$ of each slot was 160 nm, approximately a quarter of the wavelength of the light source. This ensures the resonant coupling of the incident wave to the single symmetric plasmonic mode of each slot. The gap $D$ between the two slots was approximately 50 nm. To make the case of a more general two-dimensional structure, the sample was rotated by 34° clockwise, as indicated by the two white rectangular boxes in Fig. 1c, while the polarization direction of the far-field illumination was set horizontal (red arrow in Fig. 1c). The aperture diameter $\alpha$ of the NSOM probe was 150 nm, which was much smaller than the wavelength of light source ($\lambda = 637\ nm$), but three times larger than the slot gap. The probe aperture dimension is shown in Fig. 1b for a direct comparison with the dimension of the nano-slots. For each $\vec{k}^{in}$, we scanned the NSOM fiber probe around the center of the nano-slots over an area of $800 \times 400\ nm^2$ with a scanning step of 25 nm. The distance of the NSOM probe to the sample surface was maintained at approximately 1 nm throughout the set of measurements. We repeated the same measurements for 100 different $\vec{k}^{in}$, which uniformly covered the full numerical aperture (NA = 0.6) of the bottom objective lens. The measurements of the entire angular set of complex-field maps typically lasted 27 min. Figure 1c shows the representative near-field complex-field maps for each $\vec{k}^{in} = (\vec{k}_x^{in}, \vec{k}_y^{in})$ in unit of $k_0$. These individual NSOM images, each of which corresponds to a conventional NSOM image, did not reveal the detailed structures of the nano-slots because probe aperture was significantly larger than the nano-slots.

Using the set of recorded near-field complex-field maps $E_S(x, y; \vec{k}^{in})$ shown in Fig. 1c, we constructed a FNTM, $t(x, y; \vec{k}^{in})$, by assigning 100 complex-field maps as constituent columns. It is worth noting that maintenance of the phase stability is crucial for a proper acquisition of near-field eigenmodes as the measurement is performed on a point-by-point basis. As all of the phase measurements were carried out with respect to the normal illumination whose relative phase was well controlled by the SLM, our measurements were sufficiently robust to link multiple measurements all together in their phases. Therefore, the measured FNTM was fully phase-referenced. Figure 2a shows the phase part of the FNTM. To identify the near-field modes, we performed SVD of the matrix, i.e., $t(x, y; \vec{k}^i) = U\tau V^\dagger$, where † indicates a conjugate transpose[7] and $\tau$ is a diagonal matrix whose diagonal elements are non-negative real numbers referred to as singular values. The squares of these singular values are the eigenvalues of the $t^\dagger t$ matrix. We sorted the eigenvalues in the descending order with respect to the eigenchannel index and show them in Fig. 2b. We found that the first six largest eigenvalues were above the noise level; their eigenchannels were related to the near-field eigenmodes of the double nano-slots. $V$ and $U$ are the unitary matrices whose columns are the input and output eigenchannels, respectively, associated with the corresponding eigenvalues. Therefore, the columns of $U$ contain near-field eigenmodes at the upper plane of the nano-slots.

In Fig. 2c, we visualized six near-field eigenmodes obtained from the first six columns of the unitary matrix $U$. The first column of $U$, associated with the largest eigenvalue, corresponds to the symmetric mode ($TE_{00}$ mode). The fourth column of $U$ corresponds to the antisymmetric transverse electric mode ($TE_{10}$ mode), according to the spatial phase distribution and sharp dark line in the middle. It is noteworthy that the eigenvalue of $TE_{10}$ mode was 13.4 times smaller than that of the symmetric mode ($TE_{00}$ mode) as shown in Fig. 2b. This explains why it was not visible in the individual near-field maps in Fig. 1c. Based on the shape of this antisymmetric mode, we could estimate the position and the rotation angle of the double nano-slots. The long axis of the slot was rotated by 34° with respect to the polarization of the incident wave, which agrees well with our experimental preparation. The two white rectangular boxes in the $TE_{00}$ mode map indicate the positions and orientations of the slots, which are thus identified. In a separate measurement where we rotated the sample to a different angle, the $TE_{10}$ mode was rotated accordingly. In addition, we identified other high-order modes, $TE_{01}$, $TE_{02}$, and $TE_{03}$ modes, along the direction parallel to the long axis of the slots, which correspond to second, third and fifth largest eigenvalues, respectively. These eigenmodes were excited by the polarization component of the illumination along the long axes of the nano-slots. We could identify even the quadrupole mode ($TE_{11}$ mode) whose near-field map shows the complex-field distributions of $(+,-)$ and $(-,+)$ for the two slots. In other words, the directions of the dipole moments at the two slots were opposite. This mode was rarely observed in the conventional near-field imaging as the coupling efficiency of the far-field energy to this mode was extremely low, which was manifested by its extremely small eigenvalue (28.4 times smaller than the eigenvalue of $TE_{00}$ mode). All these high-order modes reveal the structural details of the nanostructures that are smoothened out in the symmetric mode as well as the local subaperture-scale near-field waves induced by the coupling of light to the nanostructures.

The identification of the antisymmetric mode by the SVD of the measured FNTM can be understood by a simple double-slot model (see Methods for details). As the slot separation is too small for the far-field illumination, i.e., the slot gap $D$ is much smaller than the wavelength, the symmetric and antisymmetric modes cannot be individually addressed by the far-field excitation. Instead, the linear combination of these orthogonal modes was measured in the experiment, and SVD served as the means to identify individual orthogonal modes from the superposed measurements. For a given far-field incident wavevector $\vec{k}^{in}$, the phase difference of incident wave between the two slots is given by $\Delta\varphi(\vec{k}^{in}) = |\vec{k}^{in} \cdot \vec{D}|$, where $\vec{D}$ is a vector connecting the centers of the two slots. As $|\vec{k}^{in}| \leq k_0$, $\Delta\varphi \leq \frac{2\pi D}{\lambda}$ ~0.5, much smaller than $\pi$. Therefore, the incident wave is coupled mostly to the symmetric mode even at the maximum incidence angle. This explains the results that only the symmetric mode was visible in the conventional NSOM imaging (Fig. 1c) and that the eigenvalues of the higher-order modes were tens of times smaller than that of symmetric mode (Fig. 2b). In the simple double-slot model, we construct a FNTM after accounting for this far-field limit and analytically prove that the SVD can separately identify symmetric and antisymmetric modes. The eigenvalue ratio between the antisymmetric and symmetric modes is also calculated, which is given as $\frac{1}{16}\Delta\varphi^2$ in the weak-coupling regime. This model explains the experimentally measured ratio, which was on the order of $10^{-1}$-$10^{-2}$. A more general model incorporating the spatial shapes of the near-field modes was also developed to confirm the capability of our FNTM approach to extract high-order near-field eigenmodes.

As the eigenvalues of the higher-order modes were extremely small in the far-field excitation, it was crucial to increase the sensitivity of the measurements. This is especially the case given the noisy nature of the near-field recording due to weak signal strength. A large number of $\vec{k}^{in}$ measurements played an important role in this respect. Although it is sufficient for the number of $\vec{k}^{in}$'s to be equal to the number of orthogonal modes in the noise-free measurements, we used 100 different $\vec{k}^{in}$ values for the matrix measurement, far larger than the required number of measurements. This increase of the number of independent measurements raised the fidelity of the mode mapping, thereby enhancing the signal to noise ratio, particularly for the mapping of the higher-order modes. To check this out, we constructed multiple FNTMs by varying the number $N_{in}$ of $\vec{k}^{in}$ measurements used for the matrix construction out of the original 100 different measurements. By performing SVD of each matrix with $N_{in}$ columns, we identified the observable near-field eigenmodes. As shown in Fig. 2c, the FNTM with $N_{in} = 2$ revealed only the $TE_{00}$ mode because the eigenvalue of $TE_{01}$ was smaller than the experimental noise level. By contrast, the FNTM with $N_{in} = 10$ columns revealed the $TE_{01}$ near-field mode. When we increased $N_{in}$, various higher-order near-field modes were observed as the eigenvalues of the corresponding eigenmodes were increased beyond the noise level. Observing the $TE_{02}$ mode extracted in the $N_{in}$ range of 10 to 100, we could recognize that the image quality was improved with the increase in $N_{in}$. This supports the claim that a large number of $\vec{k}^{in}$ measurements played a crucial role in identifying the near-field eigenmodes. We validated the experimentally observed near-field eigenmodes by the numerical simulation using finite-difference time-domain (FDTD) method, as shown at the bottom row of Fig. 2c.

The identification of the antisymmetric mode provides an unusual opportunity to resolve the nano-slots beyond the limit set by the physical size of the NSOM probe aperture. To investigate the resolving power of the near-field eigenmode mapping method, we fabricated double and triple nano-slots whose gap was 50 nm, much smaller than the probe aperture. The direction of polarization of the incident wave was set orthogonal to the long axes of the slots. The FNTM was measured by scanning the NSOM probe through the centers of the slots, indicated by the yellow dashed line in Fig. 3a. In the experiments with the double nano-slots, the diameter of the probe aperture was 150 nm, and the probe scanning step was 25 nm. The upper row in Figs. 3a-c show a scanning electron image of the double nano-slots, the amplitude and phase profiles of the transverse near-field modes obtained by the experimentally measured FNTM, respectively. The green vertical bars indicate the positions of the nano-slots. The symmetric mode ($TE_{00}$ mode, blue curves) did not reveal the existence of the two slots as they were driven in phase. On the contrary, the antisymmetric mode ($TE_{10}$ mode, red curves) clearly resolved the two slots. The position where the destructive interference of the near-field modes between the two nano-slots occurred was distinct from the sharp amplitude dip in Fig. 3b and steep phase jump in Fig. 3c. This position exactly matches with the center of the two slots, which is a clear evidence that the identification of the antisymmetric mode greatly enhanced the resolving power. To support the experimental data, we performed numerical simulations using the FDTD method for the same configurations as in the experiment. The bottom row in Figs. 3a-c show the sample geometry used in the FDTD simulations, the amplitude and phase profiles of the near-field eigenmodes obtained by the FNTM calculated by the FDTD simulations, respectively. The excellent agreement between the experiments and simulation results supports the validity of our experiments.

The identification of the antisymmetric mode enabled us to locate the center position between the two nano-slots, which is similar to finding the center of the point-spread-function in far-field imaging. In this respect, this may not be sufficiently general to claim the resolving power. Therefore, we considered imaging of triple nano-slots to evaluate whether the proposed method can resolve the two neighboring center positions of the three nano-slots. In the experiments with the triple nano-slots, the diameter of the probe aperture diameter was 100 nm, and the probe scanning step was 15 nm. Similar to Figs. 3b and 3c, Figs. 3e and 3f show the transverse near-field modes of the triple slots. The blue and red curves represent the amplitude and phase profiles of the symmetric and $TE_{20}$ modes, respectively. Similar to the double nano-slots, the steep phase jumps of the $TE_{20}$ mode were located at the centers of the two neighboring slots because the phase difference between the adjacent slots was close to $\pi$. This led to resolving triple nano-slots whose gap size was two times smaller than the aperture diameter. These results further show the excellent agreement with the near-field modes obtained by the FDTD simulation, as shown at the bottom row of Figs. 3d-f. The theoretical limit of the resolving power is determined by the diameter of probe aperture and sensitivity of the NSOM system, and the finite scanning step of NSOM probe and the phase difference between the neighboring nano-slots are additional factors affecting to the resolving power. According to our analysis, the smallest gap that our system with 150 nm probe aperture can resolve in theory is 20 nm, but this was out of reach due to the fabrication limit of sample preparation. Further improvement of the sensitivity of the system and the adoption of advanced fabrication technology are expected to enhance the resolving power.

In conclusion, the recording of the fully phase-referenced far- to near-field transmission matrix of the subwavelength nanostructure enabled us to extract its high-order near-field eigenmodes inaccessible to the conventional near-field microscopy. As high-order modes are driven by locally varying phases of excitation, they exhibit multiple subaperture nodes due to the destructive interference of local near-field waves. This provided a new opportunity to resolve fine structural details of nano-slots whose gap was three times smaller than the physical size of the probe aperture. Considering that the steep reduction in near-field collection efficiency accompanied by the use of the smaller probe sets the practical limit of the spatial resolving power, the capability of imaging with the same resolution by the use of a larger aperture will push the ultimate limit of the resolving power in the near-field imaging. Furthermore, our approach can be combined with other existing NSOM modalities, particularly those based on interferometric detection, and can help extract near-field eigenmodes information of various quantities such as electric near-field vector components, magnetic near-field, and time/frequency-resolved measurements[13,37]. This could provide new insights for the development of novel nanoscale photonic devices.

**Methods**
**Experimental setup:** To perform the experimental mapping of near-field modes, we integrated a far-field phase modulation system into an NSOM system (Nanonics MV2000). The output beam of a laser diode (Thorlabs Inc., LP637-SF70) was enlarged and collimated to uniformly illuminate the SLM (Hamamatsu LCOS-SLM X10468). To measure both amplitude and phase of the near-field waves on the surface of the nano-slots with their phases fully referenced, we employed a self-interference measurement system without an additional setup for the reference beam line. For this purpose, the SLM generated both the sample and reference beams whose relative phase was controlled in multiples of $\pi/2$. They were de-magnified and delivered to the bottom of the NSOM sample stage using an objective lens (Nikon ELWD 40×, NA: 0.6). The overall magnification from the SLM plane to the sample stage was 1/1000×. Near-field waves generated on the upper surface of the nanostructure were measured by scanning the NSOM fiber probe. The aperture diameter of the NSOM probe was 150 nm or 100 nm. The near-field light captured by the NSOM probe was delivered to a photomultiplier tube (Hamamatsu, H8259-01).

**Simple double-slot model:** The identification of the antisymmetric mode by SVD of the measured FNTM can be understood by a simple double-slot model. The transmission of the incident field through a simple double-slot system can be described by

$$\begin{pmatrix} E_1^o \\ E_2^o \end{pmatrix} = \begin{pmatrix} J & K \\ K & J \end{pmatrix} \begin{pmatrix} E_1^{in} \\ E_2^{in} \end{pmatrix}, \tag{3}$$

where, the vectors $(E_1^{in}, E_2^{in})$ and $(E_1^o, E_2^o)$ represent the electric fields at the input and output planes, respectively (the subscripts 1 and 2 stand for the left- and right-hand slots, respectively), $J$ is the coupling constant to the output of the same slot as the input, and $K$ is the coupling to the other slot's output. In the experiment, we cannot directly measure $J$ and $K$ by the far-field excitation because the slot separation is too small for the far-field illumination to couple light to individual slots. Likewise, the symmetric and antisymmetric modes cannot be individually addressed by the far-field excitation. Instead, the combination of orthogonal modes is measured in the experiment, and SVD is used to identify the orthogonal modes from the superposed measurements. For the given far-field incident wavevector $\vec{k}^{in}$, the phase difference of the incident wave between the two slots is $\Delta\phi(\vec{k}^{in})$, as defined in the main text. The incident electric field at the two slots can be expressed as $(E_1^{in}, E_2^{in}) = (E_0, E_0 e^{-i\Delta\phi(\vec{k}^{in})})$ for any given $\vec{k}^{in}$. We sent 100 different incident wavevectors in the experiment; however, for simplicity, we consider two representative incident wavevectors, one with normal illumination and the other with $\vec{k}^{in}$. The incident electric fields at the two slots are respectively written as $(E_0, E_0)$ and $(E_0, E_0 e^{-i\Delta\phi(\vec{k}^{in})})$. Their corresponding output electric fields, which correspond to the quantities measured by the interferometric NSOM in the experiment, can be obtained by inserting these vectors into Eq. (3). Using these two output fields, we can construct a FNTM and analytically calculate its eigenvalues and eigenmodes. We could verify that the two output eigenvectors are $\vec{v}_S \approx \frac{1}{\sqrt{2}}(1,1)$ (symmetric mode) and $\vec{v}_A \approx \frac{1}{\sqrt{2}}(1,-1)$ (antisymmetric mode). Their eigenvalue ratio is calculated to be $\frac{\sigma_A}{\sigma_S} = \frac{(1-K/J)^2}{16(1+K/J)^2}\Delta\phi^2$, which is approximately $\frac{\sigma_A}{\sigma_S} \approx \frac{1}{16}\Delta\phi^2$ in the weak-coupling regime.

**Fabrication of nano-slots:** To fabricate the nano-slots with sub-50-nm spacing, we employed proximal milling techniques in $Ga^+$-based focused ion-beam processes on a sputtered gold film prepared on a silica coverslip. We intentionally off-designed the milling patterns from the original nano-slot design to make use of the proximity effect of FIB milling. By optimally controlling the distance between rectangular milling patterns and the total milling time, we could realize sub-50-nm spacing between the nano-slots.

**Acknowledgements**

This research was supported by IBS-R023-D1. Y.-H. Jin and M.-K. Kim acknowledge support received from the KBSI Project (C39212), the KIST Institutional Program (2E26680-16-P024), and the KU-KIST School Project.

**Author contributions**

E.S., Q-H.P, M.-K.K., and Wonshik C. conceived the project. E.S. carried out the measurements using the samples prepared by Y.-H.J and M.-K.K. The experimental data were analyzed by E.S., Y.-H.J., M.-K.K., and Wonshik C. E.S. and Wonshik C. developed the theoretical framework with the help of S.L. and Q-H.P. E.S., Wonjun C., Y.J., Y.-H.J., and M.-K.K. carried out the numerical simulations to support the experimental results with the help of K.-D.S. and J.A. E.S., Y.-H.J., M.-K.K., and Wonshik C. prepared the manuscript, and all authors contributed to finalizing the manuscript.

**Materials & Correspondence**

Correspondence and requests for materials should be addressed to rokmk@korea.ac.kr and wonshik@korea.ac.kr.

**Figures**

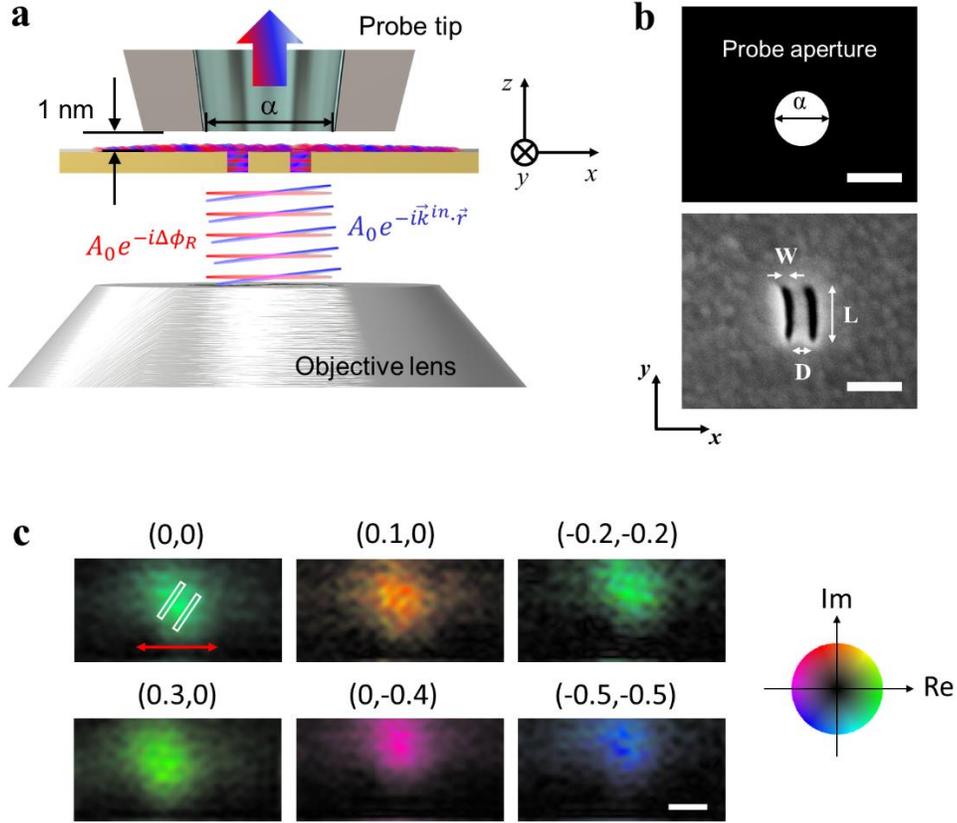

**Figure 1. Experimental recording of a far- to near-field transmission matrix. a,** Simplified experimental schematic diagram. Two planar waves, $A_0 e^{-i\Delta\phi_R}$ (red lines) and $A_0 e^{-i\vec{k}^{in}\cdot\vec{r}}$ (blue lines), generated by the SLM (not shown) were sent through the objective lens to the bottom of the nano-slots. An aperture probe made of a tapered fiber coated by a gold layer was positioned close to the upper surface. The diameter of the aperture $\alpha$ was 150 nm. The probe converted the interfered near-field wave to the far-field wave, which was then delivered to the photodetector (not shown). **b,** Upper image: the dimension of the probe aperture used in the experiment; lower image: scanning electron micrograph of the double nano-slots. $W$ and $L$ are the width and length of the individual slots, respectively, and $D$ is the gap between the two slots. Scale bar, 150 nm. **c,** Complex-field maps of the near-field waves acquired by the NSOM for various incident wavevectors, $\vec{k}^{in}$. The coordinate above each sub-figure indicates $\vec{k}^{in} = (\vec{k}_x^{in}, \vec{k}_y^{in})$ in unit of $k_0$. The red arrow indicates the polarization of the far-field illumination. Scale bar, 150 nm. Circular color map: real and imaginary values of the near-field wave. The two white rectangles indicate the boundaries of the nano-slots.

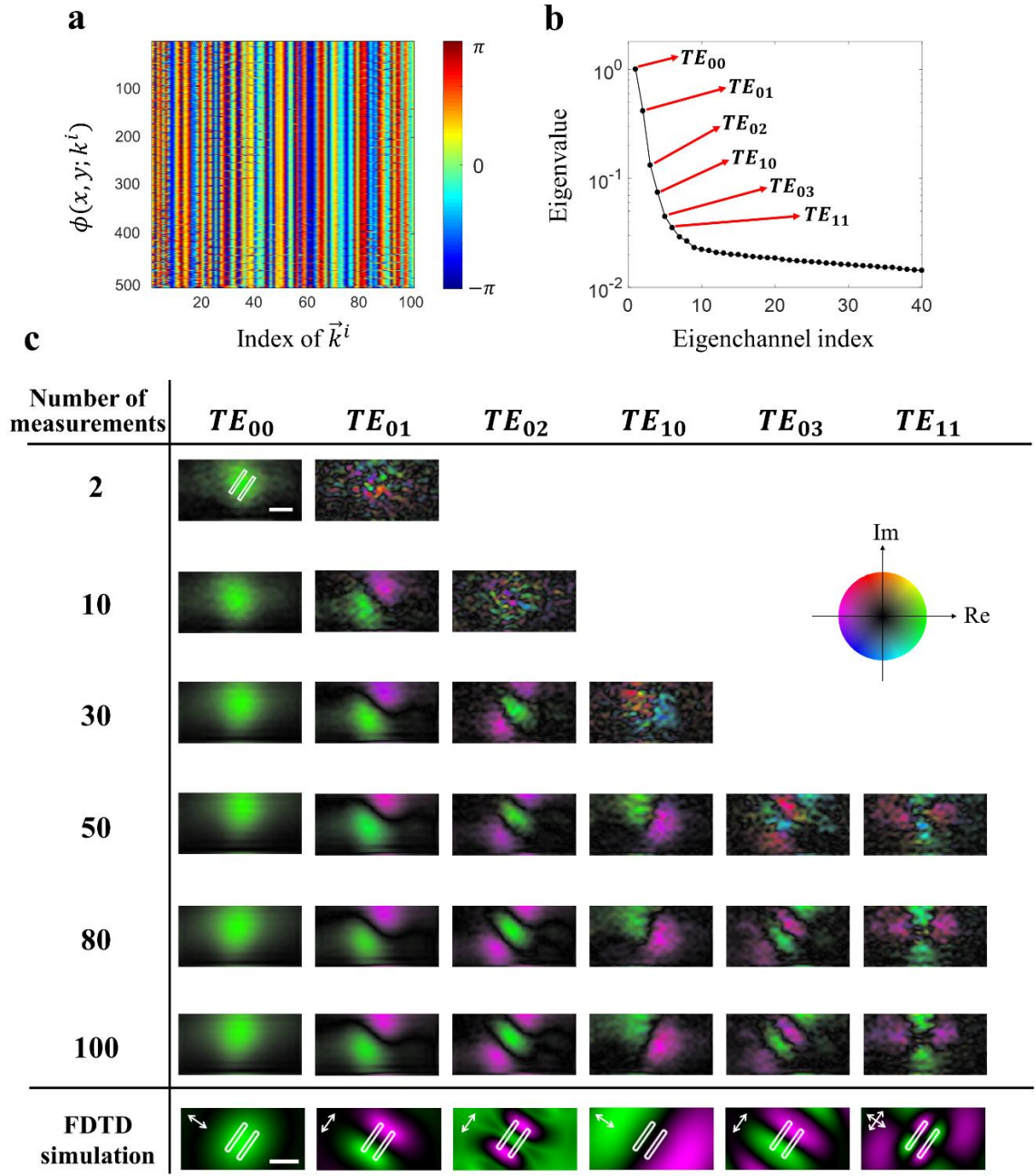

**Figure 2. Mapping of the near-field eigenmodes of the double nano-slots. a,** FNTM constructed by the near-field complex-field maps in Fig. 1c. The column index indicates $\vec{k}^{in}$ sorted in the increasing order of its magnitude. The row index describes (*x*, *y*) sorted in the increasing order of *x* and *y*. Only the phase part of the FNTM is shown. **b,** Eigenvalues of the transmission matrix sorted

in the descending order after normalizing them by the largest eigenvalue. **c,** Complex-field maps of output eigenchannels obtained from FNTMs. Among the 100 different $\vec{k}^{in}$ measurements in total, $N_{in}$ = 2, 10, 30, 50, 80, and 100 measurements were used to construct FNTMs; the corresponding near-field eigenmodes are shown. Near-field eigenmodes identified by the mode solver are shown at the bottom for comparison. Scale bar, 150 nm. Circular color map: real and imaginary values of the complex field. The white rectangles outline the boundaries of the nano-slots.

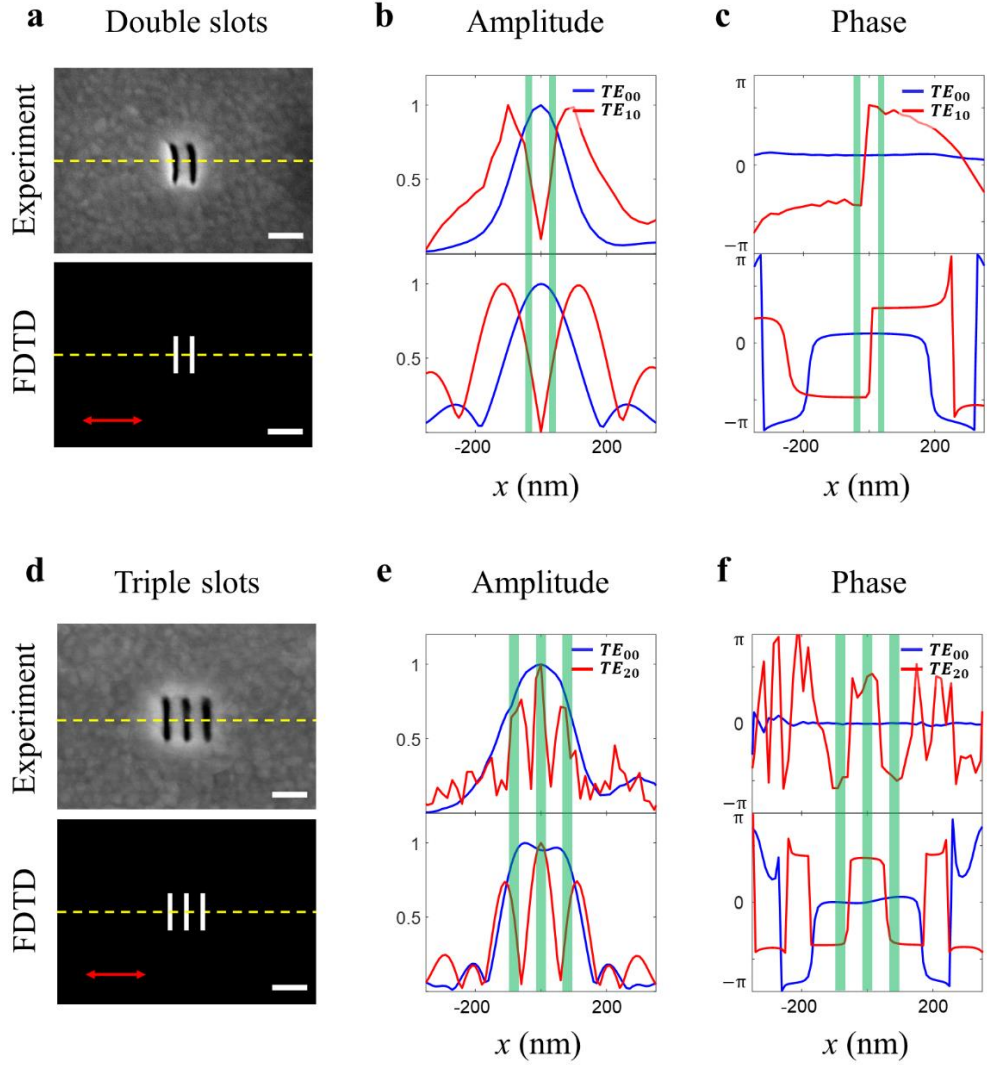

**Figure 3. Resolving multiple slots whose gap is smaller than the aperture diameter. a-c,** Near-field mode mapping of the double nano-slots with a gap of 50 nm and width of 20 nm. (**a**) Scanning electron micrograph of the sample, (**b**) amplitude profiles and (**c**) phase profiles of the eigenmodes along the yellow dashed line in **a** are shown in the upper row. The vertical green bars indicate the positions and widths of the nano-slots. The FDTD simulation results are shown at the bottom row for comparison. The red arrow indicates the polarization of the far-field illumination. Scale bar, 150 nm. **d-f,** Same as **a-c**, but for the triple nano-slots with a gap of 50 nm gap and width of 30 nm. To resolve the triple nano-slots, we obtained the $TE_{20}$ mode represented by the red curves instead of the $TE_{10}$ mode.